# Magnetic Fields Facilitate DNA-Mediated Charge Transport.

Jiun Ru Wong, Kee Jin Lee, Jian-Jun Shu, and Fangwei Shao*.

Division of Chemistry and Biological Chemistry, School of Physical and Mathematical Sciences, Nanyang Technological University, 21 Nanyang Link, Singapore 637371 (Singapore)

Exaggerate radical-induced DNA damage under magnetic fields is of great concerns to medical biosafety and to biomolecular device based upon DNA electronic conductivity. In this report, the effect of applying an external magnetic field (MF) on DNA-mediated charge transport (CT) was investigated by studying guanine oxidation by a kinetics trap ($^{8CP}$G) via photoirradiation of anthraquinone (AQ) in the presence of an external MF. Positive enhancement in CT efficiencies was observed in both the proximal and distal $^{8CP}$G after applying a static MF of 300 mT. MF assisted CT has shown sensitivities to magnetic field strength, duplex structures, and the integrity of base pair stacking. MF effects on spin evolution of charge injection upon AQ irradiation and alignment of base pairs to CT-active conformation during radical propagation were proposed to be the two major factors that MF attributed to facilitate DNA-mediated CT. Herein, our results suggested that the electronic conductivity of duplex DNA can be enhanced by applying an external MF. MF effects on DNA-mediated CT may offer a new avenue for designing DNA-based electronic device, and unraveled MF effects on redox and radical relevant biological processes.

Introduction

Electronic coupling of the highly organized array of aromatic bases down the double helical DNA makes DNA a promising biomaterial for the conduction of electrical charges, a process termed as DNA-mediated charge transport (CT).[1] Efficient DNA CT over at least 200 Å of a well-stacked duplex were readily observed.[2] Variety of approaches, from time-resolved spectroscopic measurements,[3] biochemical assays[2a,2c,4] to electrochemical methods[5] indicated that the integrity of the base pair stacking is a crucial factor in modulating CT process. Dynamic conformation of a stack of 4~5 base pairs was proposed to be the gating factor for electron hopping between adjacent base pair domains.[6] Both biological significance and technological ramifications lie in the high sensitivity to base pair integrity. DNA damage/repair and signaling proteins may harvest DNA CT as a redox-based method in vivo for fast allocation to close vicinity of genomic anomalies,[7] while molecular electronics or biosensing nanoapparatus analyze mutagenesis[5b,8] and protein-nucleic acids interactions[9] via electrochemical observation of DNA CT.

The possible negative effect of magnetic field (MF) on the genomic stability had long raised health concerns. With rapid development and wide application of magnetic based medical instruments for diagnosis and therapeutics, these concerns were further extended to the medical practice and healthcare. Despite having some progress made in theoretical and experimental studies, efforts to provide conclusive evidences to link the effect of external MF to the biological system remained a contentious issue, as there was a lack of consistent pattern in the MF exposure induced changes or damages to the cellular DNA.[10] While the underlying mechanism responsible for the biological system exerted by the external MF remains controversial, several in vitro magnetic field effects (MFE) can be taken note of. Exposure to static MF alone induces no detectable lethal effects on the cell viability and/or DNA damage regardless of the MF strength.[11] However, MF along with other stimuli, such as biological oxidants, reactive oxygen species (ROS) or ionizing radiation, could cause enhanced oxidative damage to the cellular DNA. The change in the responsiveness to MF in presence of the oxidative stress led to the postulation that MF manifests, rather than induces oxidative consequences of ROS radicals. This was not surprising as it was well documented that external MF has an extensive influence on the course and kinetics of chemical reaction containing radicals (pairs).[12] In addition, alignment of magnetically anisotropic biomacromolecules under high MF density has also been reported to be an efficient approach to form DNA films with define 3D orientations.[13] Recent conductive AFM experiments from Naaman's group also reported interesting manipulation of electron flow through monolayer of biomacromolecules via permanent magnets,[14] suggesting that MF may even affect immobilized duplex DNA on electrode surface.

Though it would provide essential fundamental knowledge to unravel the biological roles and expand technological ramification of DNA-mediated CT, the effects of external MF on charge propagation through DNA duplex and in turn how MFE affects CT-promoted oxidative DNA damage have not been well understood. Herein, we had designed a series of DNA systems to investigate how external MF can affect the electronic properties of nucleic acids and alter the yield of radical intermediates formed during DNA-mediated CT. Chemical decomposition of a kinetic fast hole trap, 8-cyclopropyldeoxyguanosine ($^{8CP}$G) after the excitation of a distant attached photooxidant, anthraquinone (AQ), is used to determine the efficiency of CT process under the absence and presence of an applied MF. Here, we observed that oxidative damages of 8CPG mediated by DNA CT through well matched and mismatched DNA are significantly elevated.

## Results and Discussions

a)
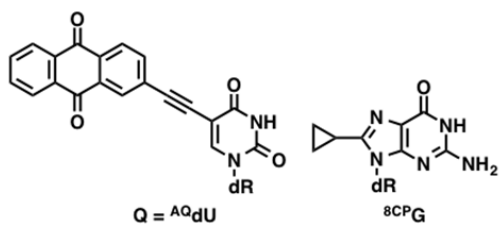

b)
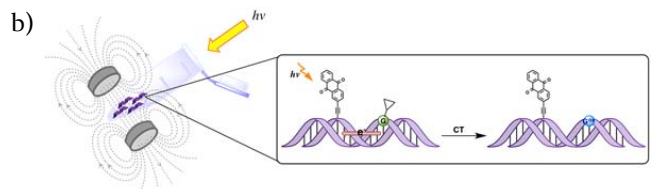

Figure 1. (a) (Top) DNA assemblies used in the present study. (Bottom) Structure of anthraquinone derivative (Q) and hole trap ($^{8CP}$G) utilized. (b) Schematic illustration of DNA-mediated CT under the presence of an external MF. A MF generated from two neodymium magnets was applied simultaneously during the excitation of the photooxidant. An electron hole is injected into the DNA duplex and is eventually trapped by a hole trap to form oxidative damage.

**Experimental Design.** Anthraquinone (AQ) is selected as the photooxidant to trigger efficient charge transport into the DNA. It had been well documented that upon photoexcitation, the AQ moiety is able to undergo rapid intersystem crossing to generate long-lived triplet states that were capable of oxidizing the natural nucleobases.[15] The attachment of AQ to deoxyuridine via an acetylene linker would ensure strong electronic coupling with the DNA π-stack and restrict the electronic injection at the anchoring site. **AQ1** and **AQ2** contain two guanine doublets, proximal and distal to the photooxidant, respectively. 5'-G of the proximal GG doublet in **AQ1** and that of the distal GG in **AQ2** are replaced with $^{8CP}$G to report the formation of guanine radical cation at corresponding GG sites via DNA CT (Figure 1a). Upon annealing to either complementary DNA or RNA strands, **DD** and **DR**, B-form DNA or A-form hybrid duplexes can be obtained. Upon irradiation at 350 nm, the excited AQ would be competent to abstract an electron from deoxyuracil and inject an electron hole into the DNA π-stack. The resulting radical cation would propagate along the base pair stacks until it is trapped by $^{8CP}$G via a rapid ring opening reaction to form permanent oxidative product (Figure 1b). DNA-CT yields can then be revealed by $^{8CP}$G decomposition via HPLC analysis after the whole duplex was enzymatically digested to nucleosides. DNA CT under an external magnetic field (MF) was performed by placing the samples between a pair of permanent neodymium magnets, which provide a highly homogenous magnetic flux across the aqueous samples (Figure 1b).

Figure 2. % Decomposition of $^{8CP}$G for B-form DNA duplexes (**AQ1-DD**, **AQ2-DD**) and A-form DNA/RNA hybrid duplexes (**AQ1-DR**, **AQ2-DR**) in 20 mM sodium phosphate buffer (pH 7.0) after irradiation for 10 min at 350 nm under the absence (crossed) and presence (shaded) of an external MF.

**Magnetic field effects accelerate $^{8CP}$G decomposition via DNA CT.** We first elucidate the influence of an applied external MF on DNA-mediated CT by monitoring $^{8CP}$G decomposition in DNA duplexes, **AQ1-DD** and **AQ2-DD** (Figure 2). In the absence of an applied MF (*B*= 0 mT), 7 % and 12 % of $^{8CP}$G at the proximal (**AQ1-DD**) and distal (**AQ2-DD**) G doublets were decomposed after 10 minutes' irradiation, respectively. Under external MF ($B$= 300 mT), both $^{8CP}$G decompositions were remarkably elevated to 33% in **AQ1-DD** and 44% in **AQ2-DD**, respectively. Slightly higher increment was observed over longer propagation length in **AQ2-DD** (32%) than that in **AQ1-DD** (26%). Irradiation of either **GG2-DD**, duplex without photooxidant, or a mixture of **AQ0-DD** and **GG2-DD**, in which AQ and $^{8CP}$G are placed in separate duplexes (Fig S1), were conducted under both $B$= 0 mT and $B$= 300 mT. No decomposition of $^{8CP}$G was observed above the noise level in both control experiments. The fact that $^{8CP}$G remained intact in the absence of AQ, regardless of external MF, indicated that $^{8CP}$G is decomposed only by photoinduced oxidation from distant AQ, and MF does not cause non-redox damage to guanines. In the second control, AQ in **AQ0-DD** does not induce any $^{8CP}$G decomposition in the separate duplex, **GG2-DD**, even in the presence of external MF. This suggested that no diffusible oxidants, such as ROS, were causing damage to $^{8CP}$G under experimental conditions either with or without the presence of MF. Thus the effects of external MF on stabilizing diffusible radical oxidants and consequently enhance guanine damage were invalid here. The elevation of $^{8CP}$G decomposition in AQ-DD duplexes after the application of an external MF was the sole consequence of MF effects on DNA-mediated charge transfer.

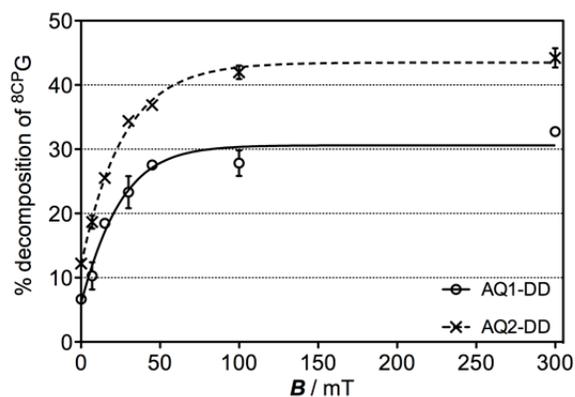

Figure 3. % Decomposition of $^{8CP}$G as a function of magnetic flux density ($B$) for **AQ1-DD** (×) and **AQ2-DD** (o) in 20 mM sodium phosphate buffer (pH 7.0) after irradiation for 10 min at 350 nm.

**Dependence of $^{8CP}$G decomposition on magnetic flux intensity**. The dependence of $^{8CP}$G decomposition on varying magnetic flux intensity ($B$= 0-300 mT) upon irradiation was further investigated and the results are showed in figure 3. It appeared that the increment of $^{8CP}$G decomposition was dependent on the strength of the magnetic flux density. Obvious MF effects on accelerating DNA CT could be detected at $B$ as low as 7 mT. CT yields, in form of $^{8CP}$G decomposition, increased linearly initially and approached plateau saturation after 100 mT for both duplexes, **AQ1-DD** and **AQ2-DD**. Throughout the entire flux intensity range, the elevation of $^{8CP}$G decomposition efficiency at distal G doublets was always more pronounced than the proximal site. Better appreciation of flux intensity increment over longer DNA bridge in **AQ2-DD** further confirmed that enhancement of guanine damage as $^{8CP}$G decomposition is due to the effects of external MF on oxidative DNA CT.

**Structural dependence**. It is well known that CT in DNA exhibits different efficiencies in various secondary structures of nucleic acids as measured by guanine oxidation,[16] fluorescence quenching,[17] electrochemistry[18] and transient absorption spectroscopy.[19] Among the various helical structures that DNA can adopt, A-form DNA/RNA hybrids are of great interest as they are essential to genetic transduction.[20] Hence, we explored whether the magnetic field effect was also applicable to A-form DNA/RNA hybrid duplex. **AQ1-DR** and **AQ2-DR** were formed by annealing RNA version of complementary strands, **DR**, to **AQ1** and **AQ2** (Figure 1a). CD spectra of **AQ1-DR** and **AQ2-DR** showed a characteristic A-form structure (Fig S2). Upon irradiation, **AQ1-DR** and **AQ2-DR** showed a decomposition of 2 % and 3 %, respectively (Figure 2). A lower $^{8CP}$G decomposition observed in the A-form hybrid duplex was not unexpected, since the appreciable interstrand stacking, due to low twist and large positive tilt, in A-form helices may not be optimal to DNA CT, as compared to intrastacking in B-form duplex.[16a,21] Albeit the lower decomposition of $^{8CP}$G, under an external MF ($B$= 0 mT), **AQ1-DR** and **AQ2-DR** showed decent increments in the damage of $^{8CP}$G to 6 % and 7%, respectively. Elevation of CT yields was not as significant as those in B-form duplex, which is probably due to that hybrid helix with wider grooves and more compact structure may not be flexible and dynamic enough to be tuned towards CT-optimal conformation by external MF. Furthermore, the higher melting temperature value for the hybrid (**AQ1-DR** and **AQ2-DR**) than the canonical B-form duplexes (**AQ1-DD** and **AQ2-DD**) indicated that the lower yield obtained was not associated with a weaker structural stability (Table S1). Regardless, MFE can also be observed in A-form hybrid duplexes and was not restricted specifically to B-form helix. In view of these results, the difference in the damage enhancement for both the proximal and distal GG under the absence and presence of an external MF suggested that charge migration was aided by exposing duplexes to an external MF.

**Scheme 1. Simplified scheme showing the pathways for photoinduced DNA CT in present work.**

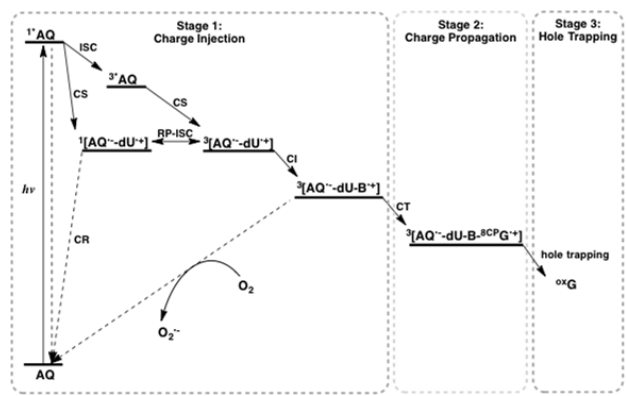

**The origins of MFE on DNA CT**. The complicated nature of a photoinduced DNA CT makes it challenging to identify the reaction steps that were affected by the external MF. Scheme 1 showed a simplified energy level diagram of three major stages of DNA CT in current system. Charge injection (CI, stage 1) was initialized upon photoirradiation of AQ to the singlet excited state, $^1$AQ*, which would rapidly undergo intersystem crossing (ISC) to yield the excited long-lived triplet state, $^3$AQ*. Subsequent charge separation (CS) between $^3$AQ* and electronically conjugated deoxyuracil would generate the initial triplet radical pair ($^3$[AQ$^{•-}$–dU$^{•+}$], $^3$RP) and a charge as a radical cation was then injected into the base pair stack. In the second stage, the radical cation migrated through bridge base pairs and would eventually oxidize $^{8CP}$G to form [$^{8CP}$G$^{•+}$]. In the final stage, $^{8CP}$G$^{•+}$, as a radical cation would undergo rapid ring opening reaction to trap the charge and complete charge transport (Scheme 1). Since it was well documented that MF could influence the spin dynamics of radical intermediates in biological reactions,[12] we postulated that external MF might affect charge injection via $^3$RP and sequential migration of radical cation, while the final stage was unlikely altered significantly by external MF.[22]

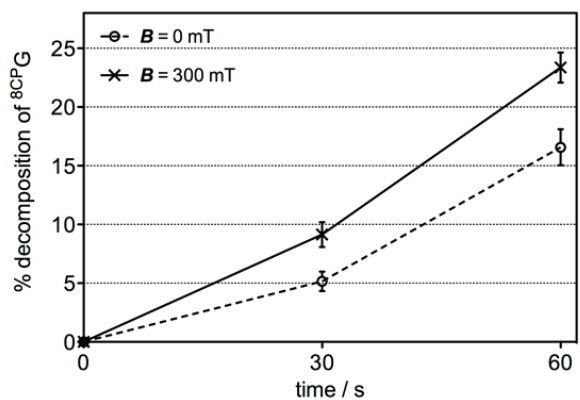

Figure 4. % Decomposition of $^{8CP}$G as a function of time in s for trinucleotides, 5'-**Q$^{8CP}$GT**-3', in 20 mM sodium phosphate buffer (pH 7.0) after irradiation at 350 nm under the absence (o) and presence (×) of an external MF (300 mT).

In the charge injection stage, the propagating radical cation on the DNA bridge, $^3$RP, ($^3$[AQ$^{•-}$–dU$^{•+}$]), may alternatively cross over to singlet radical pair ($^1$[AQ$^{•-}$–dU$^{•+}$], $^1$RP) via triplet-singlet (T–S) radical pair intersystem crossing (RP-ISC). $^1$RP would then decay to the singlet ground state by spin selective charge recombination (CR). A trinucleotide, 5'-**Q$^{8CP}$GT**-3' (Figure 1a for structure) was therefore used to elucidate whether MF can facilitate charge injection by redistributing $^3$RP and $^1$RP populations, via RP-ISC, in AQ-≡-dU system. The trinucleotide is an ideal assembly for such investigation. The flexible phosphate chain would keep photooxidant, AQ, and hole trap, $^{8CP}$G, in a close distance to ensure rapid charge injection without being limited by diffusion. The electronic coupling between AQ and $^{8CP}$G is strong enough to ensure that RP remain as a germinate pair, and dissociation of the radicals is suppressed to allow sufficient time for T–S interconversion to develop under an applied MF. Further charge migration is eliminated due to the lack of duplex formation in short trinucleotide. Figure 4 showed the decomposition of $^{8CP}$G in the trinucleotide in the absence and presence of a 300 mT magnetic flux following 0-60 s of photoirradiation. Under background magnetic field (**B**=0 mT), up to 17 % of $^{8CP}$G undergo irreversible oxidative ring-opening reaction with increasing irradiation to 60 s. Whereas under **B**= 300 mT, the trinucleotide consistently exhibit a higher decomposition (up to 23 %) during the entire irradiation time course. While our experimental setup may not be able make a distinction between the singlet and triplet radical pairs, any variation in the quantum yield of $^3$RP would directly affect the oxidative decomposition of $^{8CP}$G due to the fact that $^3$RP has a much longer lifetime than $^1$RP and should be the major species for charge injection. The appreciable enhancement in $^{8CP}$G decomposition in trinucleotide indicated that the stabilization of spin-correlated RP, specifically the triplet-state, is one of MF effects on DNA CT. Such MFE can be interpreted with reference to a simple but well-known RP model.[12] Assuming that the exchange interaction energy between the $^1$RP and $^3$RP is very small, at zero applied field, the singlet (S) and the 3 triplet sublevels (T$_0$, T$_{\pm 1}$) are nearly isoenergetic and there will be unrestricted RP-ISC between the 4 states, induced by electron-nuclear hyperfine interaction (HFI). As the applied MF strength increased, the electronic Zeeman interaction would slow down and uncoupled the T$_{\pm 1}$ from the conversion process. Eventually, only the interconversion between T$_0$–S states would be remained at high flux intensity (hyperfine mechanism). Hence the conversion of $^3$RP to $^1$RP was diminished and decay via charge recombination to the singlet state would be impeded. Consequently, the populations of $^3$RP would increase and charge injection from $^3$RP as an 'escape' product from CR would be enhanced and be revealed as more hole trapping at $^{8CP}$G site. Hence in trinucleotide model, higher efficiency of $^{8CP}$G decomposition under external MF implied that charge injection is facilitated by MF via enriching the triplet radical pair, $^3$[AQ$^{•-}$–dU$^{•+}$] during photoinitiated charge separation.

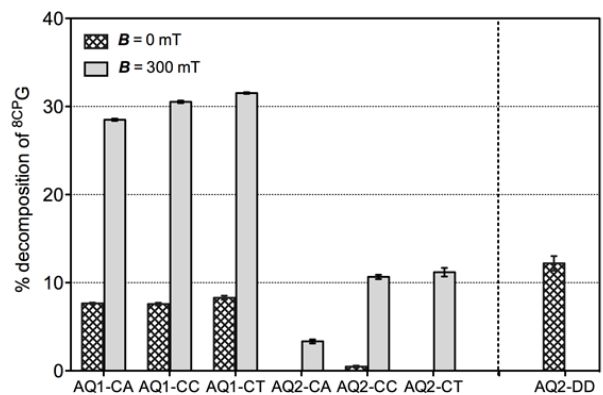

Figure 5. % Decomposition of $^{8CP}$G for **AQ1**- and **AQ2-CA**, **CC** or **CT** in 20 mM sodium phosphate buffer (pH 7.0) after irra-

diation for 10 min at 350 nm under the absence (crossed) and presence (shaded) of an external MF.

Notably, it was well accepted that MF could induce orientation in organic molecules[23] and biological macromolecules.[24] The preferential orientation originates from the external magnetic induction and the anisotropy susceptibility of the DNA nucleic bases.[23,25] Although the magnetic susceptibility of neutral nucleobases are weak, paramagnetic radical cation as the propagating species in DNA CT, could exhibit a much stronger response to MF. An intervening mismatch is a severe disruption to the integrity of DNA base pair stacking and could significantly diminish the overall CT yields.[2a,8,26] Hence we challenge the ability of MF to facilitate DNA CT by introducing different mismatches to DNA bridge before and after $^{8CP}$G trap. Upon annealing to complementary strands with substitution of G by either A, C or T, duplexes containing **CA**, **CC** and **CT** mismatches between proximal and distal GG were formed and submitted to photoinitialized DNA CT under $B$= 0 mT and $B$= 300 mT (Figure 1a). When **AQ1-CA**, **AQ1-CC** and **AQ1-CT** were excited under $B$= 0 mT, $^{8CP}$G at the proximal GG site in all three duplex showed a decomposition of ~7 %. After $B$= 300 mT was applied, $^{8CP}$G damage was increased to ~30 % (Figure 5). These results were almost identical to **AQ1-DD**. None of the mismatches diminished DNA CT between photooxidant and proximal GG, which was not unexpected as the mismatched base pair was positioned after the hole trap and the intervening base pair stack between AQ and $^{8CP}$G was not disturbed. With the photoinduced DNA CT remaining intact and similar degree of efficiency enhancement being observed, it implied that MFE was not restricted by the presence of a mismatched site in the duplex. Whereas, in the cases of **AQ2-CA**, **AQ2-CC** and **AQ2-CT**, damage yield of distal $^{8CP}$G under $B$= 0 mT was almost fully quenched by a single base interruption. Similar to previous studies on CT chemistry,[8a,8b,26] base replacement of G with A, C or T disrupted the integrity of the π-stack, and inhibited charge propagation to the distal $^{8CP}$G. This showed that CT under our DNA assembles is still sensitive to the integrity of base pair stacking as previously reported. Interestingly, under $B$= 300 mT, decomposition yields of $^{8CP}$G were recovered in **AQ2-CA** (4 %), **AQ2-CC** (11 %) and **AQ2-CT** (11 %). In the cases of two pyrimidine/pyrimidine mismatches, CT yield was restored to a similar efficiency as the well-matched DNA under background MF. The reinstallation of DNA CT by external MF suggested that MF had applied a well-pronounced compensation effect to repair or shield the distortion of base pair stacking. Characterization studies by X-ray crystallography and NMR methods had shown that mismatch base pair cause minimum alterations on the global conformation of the B-DNA duplex, but the distortions were localized in the vicinity of the mismatched site.[27] Hence, it is likely that a MF as weak as 300 mT might be sufficient to confer partial base orientation to the mismatched and the neighboring base pairs so that the optimal π-stacking in the local environment can be recovered temporarily to the CT-active conformations comparable to a well-matched duplex and sequentially efficient CT can be assessed. Though the current experiments cannot distinguish whether the compensation effects adjusted the dynamic conformation of duplex when DNA was still in neutral form or when DNA was oxidized radical and cation was transiently occupying the domain of base pairs, we tends to believe that the latter should be more significant due to low flux intensity we applied in the experiments.

The ability to accelerate CT in duplex DNA by the application of a weak MF could be the basis for designing a magnetically controlled DNA electronic switch. The magnitude of the MF-induced signal could be manipulated by changing the nature of the mismatch base pair or the DNA secondary structures. In a mismatched duplex, optimum base stacking was not achieved and charge propagation would be shut off beyond the mismatch site. Without applying MF, the mismatched duplex wire will be in an "off" state. Likewise, switching on the MF without irradiation would not create any detectable DNA-mediated CT signal. However, when two triggering fuels, an excitation source and an applied MF, were concurrently present, the switch can then be turn "on". Subsequent removal of either fuel would restore the "off" state. Coupled with electrochemical device, MF alone could also be used as a switch control for currents.

Given that DNA-mediated CT can report on the integrity of DNA over long molecular distances, it had been recently proposed that DNA-bound repair protein with a metal cluster as the redox cofactor might exploit DNA CT to signal and communicate with each other and so as to efficiently detect lesions across the genome.[7c] Interestingly, our results suggested that with the assistance of MF, local alignment near the mismatch site might permit the charge migration to proceed without obstruction. As such, electron may still be able to shuttle through the damaged DNA between the two repair proteins in the presence of a mild MF. Consequently, the proteins may detach and bind to an alternative DNA site without first locating and repairing the damage. In essence, the proteins will erroneously pass on false information about DNA structural integrity to one another through the transiently aligned DNA π-stack by MF. Potentially, this could be a reasonable justification why more DNA damages were detected in the presence of an applied MF in the biological systems. In fact, the number of lesion sites generated by ROS could remain unchanged except that in the presence of MF, the repair system was compromised since repair enzymes were misled by MF-assisted DNA CT and bypass the screening of the lesion sites, leading to an "accumulation" of damage spots.

### Conclusion

Here by using anthraquinone as photooxidant and $^{8CP}$G as hole trap, CT efficiencies through duplex DNA were explored in the presence of an external MF. The application of MF caused an increased damage to $^{8CP}$G via photoinitialized DNA CT over various lengths of duplex bridge.

Such effects were also observed in the DNA/RNA hybrid duplex, albeit lower yields. The acceleration of CT is closely related to the strength of the applied MF, where damage to $^{8CP}$G was detectable as low as 7 mT. Two factors may account for the observed MF-induced acceleration of DNA CT. Application of an external MF interfered the spin evolution of RP and consequently, enhanced the of the triplet state population and enhanced hole injections into the DNA π-stack. Secondly, MF can manipulate the alignments of base pair stacking to promote high accessibility of DNA bridge to CT-active conformation. Herein, our results suggest that external MF could be a promising approach to enhance the electronic conductivity of duplex DNA. This approach would have open a new venue to design DNA CT based molecular device and inspire a better understanding of related biological and medical processes.

**Experimental Section.**

**Oligonucleotides synthesis.** Cyanoethyl phosphoramidite of 8-cyclopropylguanosine[28] were synthesized as described while anthraquinone-5-ethynyl-dU was purchased from Berry & associates. DNA oligonucleotides with trityl-on were synthesized using standard phosphoramidite protocols on Bioautomation Mermade 4 DNA synthesizer with reagents from Glen research. After incubation in AMA at 37 ºC for 2 h, the cleaved DNA strands were purified by reverse phase HPLC (mircosorb 100-5 C18 Dynamax, 250 × 10.0 mm), detritylated wth 80% glacial acetic acid for 15 min and repurified by reverse-phase HPLC. All the DNA oligonucleotides were confirmed by ESI mass spectrometry and quantified by UV-vis spectroscopy.

**Photooxidation experiment.** DNA duplexes (10 μM, 30 μL in 20 mM sodium phosphate buffer, pH 7.0) were prepared by annealing the modified DNA strands with its complements (ratio of 1: 1.1) and gradually cooled to room temperature overnight after heating for 5 min at 90 ºC. The duplexes were then irradiated with a 450 W Xenon lamp, equipped with monochromater and a 320 nm longpass filter for 10 min. Exposure to a stationary magnetic field was performed by placing 2 neodymium magnets (0.7 cm in diameter) on each side of the mircotube that contained the samples. The opposite poles of the magnets were separated by about 0.7 cm. As the surface area of the magnetic disc was larger than that of the sample area, it was considered that the magnetic field strength of ~ 300 mT between the opposite poles was homogenous. Following, the samples were digested into free nucleosides by incubating at 37 ºC with phosphodiesterase I and alkaline phosphatase for 24 h. The nucleosides were subsequently separated and analyzed by reverse-phase HPLC (Chemcobond 5-ODS-H, 4.6 × 150 mm).

## ASSOCIATED CONTENT

**Supporting Information**. Supporting figures and table associated with this article is available free of charge via the Internet at http://pubs.acs.org."

## AUTHOR INFORMATION

**Corresponding Author**

* E-mail: fwshao@ntu.edu.sg

## ACKNOWLEDGMENT

The financial support for this research work by Ministry of Education of Singapore (M4011040, M4020163) and Nanyang Technological University (M4080531) is greatly appreciated.

## ABBREVIATIONS

CT, charge transport; MF, magnetic field; MFE, magnetic field effects; ROS, reactive oxygen species; $^{8CP}$G, 8-cyclopropyldeoxyguanosine; AQ, anthraquinone; CI; charge injection; ISC, intersystem crossing; CS, charge separation; RP, radical pair; CR, charge recombination.

Insert Table of Contents artwork here

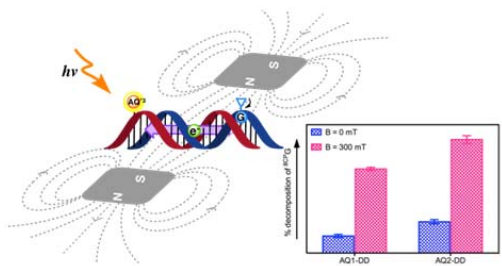